\documentclass[conference,a4paper]{IEEEtran}
\newcommand{\Z}{\mathbb{Z}}
\newcommand{\R}{\mathbb{R}}

\newcommand{\N}{\mathbb{N}}

\usepackage{amsfonts, amssymb, amsthm}
\usepackage{balance}
\usepackage{bm}
\usepackage{xcolor}
\usepackage[utf8]{inputenc} 
\usepackage[T1]{fontenc}
\usepackage{url}              
\usepackage{cite}             

\usepackage[cmex10]{amsmath}  
\usepackage{mleftright}       
\mleftright                   

\usepackage{graphicx}         
\usepackage{booktabs}         

\newtheorem{thm}{Theorem}
\newtheorem{lemma}{Lemma}

\theoremstyle{definition}
\newtheorem{defn}{Definition}
\newtheorem*{rmk}{Remark}

\DeclareMathOperator{\bern}{Bern}
\let\P\relax
\DeclareMathOperator\P{\textsf{P}}
\begin{document}
\title{Constructions for Nonadaptive\\ Tropical Group Testing}

 \author{
  \IEEEauthorblockN{Nicholas Kwan and Lele Wang}
  \IEEEauthorblockA{University of British Columbia, Vancouver, BC V6T1Z4, Canada\\
                   \texttt{nickwan@student.ubc.ca, lelewang@ece.ubc.ca}}
                  }

\maketitle


\begin{abstract}
PCR testing is an invaluable diagnostic tool that has most recently seen widespread use during the COVID-19 pandemic. A recent work by Wang, Gabrys and Vardy proposed tropical codes as a model for group PCR testing. For a known but arbitrary number of infected persons, a sufficient condition on the underlying block design of a zero-error tropical code, called double disjunction, is proposed. Despite this, the parameters for which the construction of doubly disjunct block designs is known to exist are very limited. In this paper, we define probabilistic tropical codes and consider random block designs that are doubly disjunct with high probability. We also provide a deterministic construction for a doubly disjunct block design given a disjunct block design. We show that for certain choices of parameters, our probabilistic construction has vanishing error. Our constructions, combined with existing methods, give us three different ways to construct tropical codes. We compare the number of tests required by each, and bounds on the error.
\end{abstract}

\section{Introduction}
\label{sec:Introduction}

Polymerase Chain Reaction (PCR) is a method by which genetic material may be rapidly amplified. PCR has a wide range of applications, perhaps most recently in reliable testing for the virus causing COVID-19. PCR testing is typically done through thermal cycling, where the viral load of a test containing some specimens
is repeatedly doubled in concentration.
Each doubling event is known as a cycle, and the load is doubled until it reaches a detectable threshold or fails to reach this threshold within a preset number of cycles. The cycle number at which this threshold is reached is called the \textit{Ct value}. The Ct value of a test is thus $$\text{Ct value} = \lfloor -\log_2(\text{viral load}) + \text{constant}\rfloor,$$ and is a measure of the degree of infection of a test, with smaller values indicating higher degrees of infection. If a test fails to reach the threshold after a preset number of cycles, then it is not infected and its Ct-value is $\infty$.

Consider some finite set $\mathcal{S}$ of people with an infected subset $\mathcal{D} \subset \mathcal{S}$. Group testing deals with the problem of identifying $\mathcal{D}$ by performing tests on subsets of $\mathcal{S}$ whose size may be greater than one. By nonadaptive group testing, we consider a setting where all group tests are given at once. Group testing can be divided into combinatorial and probabilistic group testing. Combinatorial group testing identifies $\mathcal{D}$ with zero error, while probabilistic group testing identifies $\mathcal{D}$ with high probability. Traditionally, group testing takes place in the binary setting so that a test on some subset $\mathcal{U} \subset \mathcal{S}$ is positive when $\mathcal{U}\cap \mathcal{D}$ is nonempty and negative otherwise. On the other hand, a PCR test on a subset $\mathcal{U}$ returns some aggregate Ct-value, which tells us more than whether or not $\mathcal{U}\cap \mathcal{D}$ is nonempty. For instance, when two specimens with Ct values $x$ and $y$ are placed into a PCR test at the same time, the resulting Ct value is $z$ satisfying 
\[
2^{-x} + 2^{-y} = 2^{-z}.
\]
Due to exponential decay, we have $z \approx \min\{x,y\}$.
In performing a group PCR test, we are also able to add a specimen to a test after a certain number of cycles have elapsed. The operation of adding a specimen to a test after some $\delta$ cycles is known as delaying, and we say that the specimen has been delayed by $\delta$ cycles. When a specimen with Ct value $x$ is delayed by $\delta$ cycles, the number of cycles it takes to reach the same detectable threshold becomes $x + \delta$.

Based on the above properties of PCR testing, Wang, Gabrys, and Vardy proposed the tropical semiring as a model for group PCR testing~\cite{Wang--Gabrys--Vardy2022}.
The tropical semiring $(\R\cup\{\infty\}, \oplus, \odot)$ consists of the real numbers and infinity equipped with the following operations:
\begin{align*}
    x \oplus y &:= \min(x, y),\\
    x \odot y &:= x+y.
\end{align*}
The additive identity is $\infty$ and the multiplicative identity is $0$. In this model, for two specimens $x$ and $y$ (we have identified the specimens with their individual Ct-values), the Ct-value $(x\odot\delta_x)\oplus(y\odot\delta_y) = \min(x+\delta_x, y+\delta_y)$ is given by a test containing both $x$ and $y$, delayed each by $\delta_x$ and $\delta_y$ respectively. In this way, $\odot$ indicates delay and $\oplus$ indicates presence. This model works because viral loads vary widely, and the Ct-value of a test with two specimens with differing viral loads is dominated by the specimen with the higher viral load due to the exponential nature of the test. Since Ct values and delays only take nonnegative integer values, all the analysis done in this paper is over the subsemiring $(\N \cup \{0, \infty\}, \oplus, \odot)$.

In general, a test involving specimens $x_1, \ldots, x_k$ with delays $\delta_1, \ldots, \delta_k$ will return the Ct-value $\bigoplus_{i=1}^k(x_i \odot \delta_i)$. In this way, we can define tests using tropical matrix multiplication. Consider a set of $n$ individuals, and let $x = [x_1, x_2, \ldots, x_n]^\intercal \in (\N \cup \{0, \infty\})^n$ be a list of their Ct values. A set of $t$ tests on these individuals can be formally written as a matrix of delay values $S \in (\N \cup \{0, \infty\})^{t \times n}$, known as the schedule matrix. If $S_{ij} = \infty$, then the $i^\text{th}$ test did not involve the $j^\text{th}$ individual. The results of the tests are then given by $S \odot x$, where here, $\odot$ denotes tropical matrix multiplication; that is, $$S \odot x = \begin{bmatrix}
    \bigoplus_{j=1}^N(S_{1j}\odot x_j)\\
    \vdots\\
    \bigoplus_{j=1}^N(S_{tj}\odot x_j)
\end{bmatrix}.$$
If, for any distinct $x, y \in (\N \cup \{0, \infty\})^n$ with at most $d$ finite values, $S\odot x \ne S \odot y$, then our schedule matrix is a valid $(t, n, d)$-tropical code in the combinatorial setting. The code is said to have maximum delay $l$, where $l$ is the largest finite element in $S$. In a real-world setting, $l$ is typically held at 40 cycles. For general $d$, zero-error tropical codes are hard to construct. 

Existing work in nonadaptive group testing considers the construction of binary matrices satisfying certain properties, such as disjunction and separability, that correspond to binary group testing schemes. These properties are necessary, but not sufficient, for binary matrices that  correspond to the schedule matrix of a tropical code. Furthermore, the sufficient doubly disjunct property given in definition~\ref{def:doublydisjunct} by~\cite{Wang--Gabrys--Vardy2022} is new, and to our knowledge, existing constructions do not exist for doubly disjunct matrices. In this paper, we propose the construction of \emph{random} schedule matrices that are doubly disjunct with high probability, as well as the construction of doubly disjunct matrices from disjunct matrices.

{\bf Notation.} Throughout we use calligraphic letters such as $\mathcal{K}$ to refer to sets and block designs. We use boldfaced letters for random matrices $\bm{S}$ and random vectors $\bm{s}$ when they are capitalized and in lowercase, respectively. $\P(A)$ refers to the probability of the event $A$, and we define $p_{X}(x) \triangleq \P(X = x)$. Plain capital letters are used for matrices and blocks, and $\vec{p}$ is used for vectors.

\section{Summary of Existing Results}
\label{sec:summary}
This paper relies heavily on the results given by~\cite{Wang--Gabrys--Vardy2022}, which was the first paper to consider using the tropical semiring as a model for group PCR testing, and produced results on tropical codes in the combinatorial setting. 

For some $t \in \mathbb{N}$, a block design on $t$ vertices is a subset $\mathcal{F} \subset 2^{[t]}$ of the power set $2^{[t]}$ satisfying certain properties with respect to unions, intersections, and differences. 
Each element $B \in \mathcal{F}$ is called a block. The set operation properties satisfied by the blocks vary between block designs.
An example of a block design on $7$ vertices is the Fano plane $\mathcal{F} = \{\{1, 2, 4\}, \{1, 3, 7\}, \{1, 5, 6\}, \{2, 3, 5\}, \{2, 6, 7\}, \{3, 4, 6\},$ $\{4, 5, 7\}\}$. Note that the intersection of any two blocks has exactly one element, and that each block has size three. Furthermore, each element of $[7]$ is contained in exactly three blocks.

We may associate a block design $\mathcal{F}$ on $T$ vertices with $|\mathcal{F}| = N$ with an incidence matrix $M$,
a $T \times N$ binary matrix where each column corresponds to a block. Then 
$$M_{ij} = \begin{cases}
0 & \text{if $i \notin B_j$ for the block $B_j \in \mathcal{F},$}\\
1 & \text{if $i \in B_j$ for the block $B_j \in \mathcal{F}.$}
\end{cases}$$ Note that this representation is unique up to permutation of the columns $M$, whereas the permutation of the rows of $M$ results in an equivalent block design. For instance, the incidence matrix of the previously described Fano plane is 
$$\begin{bmatrix}
1 & 0 & 0 & 0 & 1 & 0 & 1\\
1 & 1 & 0 & 0 & 0 & 1 & 0\\
0 & 1 & 1 & 0 & 0 & 0 & 1\\
1 & 0 & 1 & 1 & 0 & 0 & 0\\
0 & 1 & 0 & 1 & 1 & 0 & 0\\
0 & 0 & 1 & 0 & 1 & 1 & 0\\
0 & 0 & 0 & 1 & 0 & 1 & 1
\end{bmatrix}.$$
We may use the incidence matrices of suitable block designs to construct tropical codes by replacing each $0$ with $\infty$ and each $1$ with a suitable finite delay value. Conversely, every tropical code has an associated block design. 

\begin{defn}[Doubly Disjunct Block Design]
\label{def:doublydisjunct}
A block design $\mathcal{F}$ with $n$ blocks on $t$ vertices is said to be $d$-doubly disjunct if, for any distinct blocks $Z, B_1, \ldots, B_d \in \mathcal{F}$, $\left|Z \setminus \left(\bigcup_{i=1}^d B_i\right)\right| \ge 2.$

If instead $\left|Z \setminus \left(\bigcup_{i=1}^d B_i\right)\right| \ge 1$, then we call $\mathcal{F}$ $d$-disjunct.
\end{defn}
\begin{rmk}
Clearly if $\mathcal{F}$ is a $d$-doubly disjunct block design, it is also $d$-disjunct.
\end{rmk}
The Fano plane is a 1-doubly disjunct block design on $7$ vertices. A construction for $1$-doubly disjunct block design on some arbitrary number of vertices $t$ is given by Graham and Sloane~\cite{Graham--Sloane1980}.
Let $\mathcal{S}^t_w$ be the set of all binary vectors of length $t$ and Hamming weight $w$, consider the map 
\begin{align*}
f: \mathcal{S}_w^t &\longrightarrow \Z/t\Z,\\
\vec{v} &\longmapsto \left(\sum_{i=1}^n iv_i\right) \mod t,
\end{align*}
taking a binary vector $v$ to the sum of the indices of its ones, modulo $t$. Then the preimage of any residue class will be a $1$-doubly disjunct block design. At least one of these preimages will have cardinality larger than $\frac{1}{t}\binom{t}{w}$.

\begin{thm}[Theorem 31 in~\cite{Wang--Gabrys--Vardy2022}]
\label{thm:code from doubly disjunct}
Let $\mathcal{F}$ be a $(d-1)$-doubly-disjunct block design with $n$ blocks on $t$ vertices. Let $M$ be the incidence matrix of $\mathcal{F}$. Then there exists a $(t, n, d)$-tropical code whose associated block design is $\mathcal{F}$. That is, the finite entries of the schedule matrix $S$ coincide exactly with the nonzero entries of $M$.
\end{thm}
In the case where $d = 2$ and each block $B \in \mathcal{F}$ has size at least three, there exists a decoding algorithm, given by Theorem 17 in~\cite{Wang--Gabrys--Vardy2022}, whose maximum delay is a prime bounded above by $n$. 
Otherwise, for $d > 2$, the decoding algorithm given in Theorem 31 in~\cite{Wang--Gabrys--Vardy2022} uses a maximum delay of $2^{n(t+1)}$, with the schedule matrix
$$S_{ij} = \begin{cases}
    2^{i+jt}, & \text{ if } M_{ij} = 1,\\
    \infty, & \text{ if } M_{ij} = 0.
\end{cases}$$
although the authors conjecture that there is a polynomial delay code using the same underlying block design. The given decoding algorithm considers sets of potentially infected persons of size $d$. We may construct a bipartite graph using a plausible set of $d$ infected persons and the actual set of $d$ infected persons as our two sets of vertices. The double disjunctness of $\mathcal{F}$ guarantees that this graph contains an even cycle. The alternating sum along this cycle will correspond to the sum of delays for the actual set of infected persons.

To the best of our knowledge, there is no construction for a $(d-1)$-doubly disjunct block design for $d > 2$ with arbitrary parameters. Existing constructions for $d$-disjunct matrices, which are suitable for binary group testing, exist. Combinatorial constructions exist with $t = O(d^2\log n)$~\cite{Kautz--Singleton1964}, while Monte-Carlo constructions, like our own, exist for $ O\left(d^2\min\{\log n, (\log_t n)^2\}\right)$~\cite{Cheraghchi--Nakos2020}.
Probabilistic binary group testing schemes exist with $t = \Theta(d\log n)$~\cite{Inan--Kairouz--Wootters--Ozgur2019}.

\section{Result I: A Deterministic Construction from Disjunct Block Designs}

Recall from Definition~\ref{def:doublydisjunct} what it means for a block design $\mathcal{F}$ to be disjunct and doubly disjunct. Our first result provides a simple way to construct $d$-doubly disjunct block designs from $d$-disjunct block designs by doubling the number of vertices. These can then be used to produce $(t, n, d)$-tropical codes by Theorem~\ref{thm:code from doubly disjunct}. Methods for constructing $d$-disjunct block designs are given in~\cite{Inan--Kairouz--Wootters--Ozgur2019, Cheraghchi--Nakos2020, Kautz--Singleton1964}. 
\begin{thm}
\label{thm:disjunct to doubly disjunct}
    Let $\mathcal{F} \subset 2^{[t]}$ be a $d$-disjunct block design on \(t\) vertices. Then the image $\mathcal{G}$ of the function 
    \begin{align*}
        f : \mathcal{F} &\longrightarrow 2^{[2t]}\\
        B &\longmapsto \{2x : x \in B\}\cup\{2x-1: x \in B\},
    \end{align*}
    is a $d$-doubly disjunct block design on \(2t\) vertices.
\end{thm}
\begin{IEEEproof}
[\bf Proof]
    Let \(B_0, \ldots, B_d\) be any \(d+1\) distinct blocks in \(\mathcal{G}\), and for each $i \in \{ 0, \ldots, d\}$ let \(A_i\) be the unique element of \(\mathcal{F}\) mapping to \(B_i\). Since \(\mathcal{F}\) is disjunct, \(\left|A_0 \setminus\left( \bigcup_{i=1}^dA_i\right)\right| \ge 1\). For each block, the maps $x \mapsto 2x$ and $x \mapsto 2x-1$ are injective, and the images of the two maps are disjoint. Let \(a \in A_0 \setminus\left( \bigcup_{i=1}^dA_i\right)\). Then note that \(2a \in B_0\) and \(2a-1 \in B_0\). If $2a \in \bigcup_{i=1}^dB_i$, then \(2a \in B_i\) for some \(i \ge 1\), but this would imply that \(a \in A_i\) for some \(i \ge 1\), which cannot be the case. Similarly, if $2a-1 \in \bigcup_{i=1}^dB_i$, then \(2a-1 \in B_i\) for some \(i \ge 1\), but this would imply that \(a \in A_i\) for some \(i \ge 1\), which cannot be the case. Hence, \(2a \in B_0 \setminus\left( \bigcup_{i=1}^dB_i\right)\) and \(2a-1 \in B_0 \setminus\left( \bigcup_{i=1}^dB_i\right)\), so \(\left|B_0 \setminus\left( \bigcup_{i=1}^dB_i\right)\right| \ge 2\) and \(\mathcal{G}\) is \(d\)-doubly disjunct.
\end{IEEEproof}
\begin{rmk}
    This construction extends to the construction of $n$-fold disjunct block designs, but it is not optimal. For instance, consider the Fano plane as described in section \ref{sec:summary}. The Fano plane is an ``optimal'' 2-disjunctive block design on 7 vertices in the sense that no blocks can be added to it while preserving 2-disjunctiveness, but applying Theorem \ref{thm:disjunct to doubly disjunct} to the Fano plane and adding the block $\{2, 4, 6, 8, 10, 12, 14\}$ still yields a 2-doubly disjunct block design on 14 vertices.
\end{rmk}

\section{Result II: A Probabilistic Construction}
In this section we define tropical codes in a probabilistic setting, and construct random matrices that are doubly disjunct with high probability.
\subsection{Definition}
Consider a \emph{random} $t \times n$ schedule matrix $\mathbf{S}$. We define the probability of error as
\begin{align*}
    P_e^{n,d}(\bm{S}) \triangleq \P\left({\bm S}\odot \vec{x} = {\bm S}\odot \vec{y} \text{ for some distinct } \vec{x}, \vec{y} \in \mathcal{R}_{n,d}\right),
\end{align*}%
where 
$\mathcal{R}_{n,d} \triangleq \{\vec{x} \in (\N \cup \{0, \infty\})^n:  x \text{ contains at most $d$}\\ \text{finite values}\}$.
In other words, $P_e^{n, d}(\bm{S})$ is the probability that a group testing scheme with schedule matrix $\bm{S}$ confuses $x$ and $y$.
To formalize the definition of a tropical code in a probabilistic setting, we introduce a parameter $\epsilon$ to bound the probability of error $P_e^{n, d}(\bm{S})$.
\begin{defn}
    A random $t\times n$ schedule matrix $\bm{S}$ is a $(t, n, d, \epsilon)$-tropical code if $P_e^{n, d} \le \epsilon$.
\end{defn}
A $(t, n, d, 0)$-tropical code then corresponds to the definition of a $(t, n, d)$-tropical code given in~\cite{Wang--Gabrys--Vardy2022}. This definition naturally motivates us to find relationships between the parameters $t, n, d$ and $\epsilon$.

\subsection{Construction}
\label{sec:construction}

In this section we find parameters $(t, n, d)$ for which we are able to construct doubly disjunct matrices with high probability, then compute $\epsilon$ for our construction, based on its probability of generating a matrix that is not doubly disjunct.

We show the existence of a joint distribution $p_{\bm s}(\vec{v})$ for the first row of delays $\bm{s}$ in $\bm{S}$. We then generate the remaining rows of $\bm{S}$ i.i.d. according to the same distribution $p_{\bm s}(\vec{v})$. 

We produce sufficient properties on $p_{\bm s}(\vec{v})$ for $\bm{S}$ to be doubly disjunct with high probability.
Since $p_{\bm{s}}(\vec{v})$ is a probability distribution, we require
\begin{equation}
\label{cond: sum to one}
    \sum_{\vec{v} \in \{0, 1\}^n}p_{\bm{s}}(\vec{v}) = 1,
\end{equation}
and 
\begin{equation}
\label{cond: all positive}
    p_{\bm{s}}(\vec{v}) \ge 0 \text{ for all } \vec{v} \in \{0, 1\}^n.
\end{equation}
For $\bm{S}$ to be $d-1$-doubly disjunct on average, we require, for every distinct $i_1, i_2, \ldots, i_d \in [n]$
\begin{align}
\label{cond: dd whp}
\P\begin{pmatrix}
\bm{s}_{i_1}=1 \text{ and}\\
\bm{s}_{i_2} = \cdots = \bm{s}_{i_d} = 0 
\end{pmatrix}
=\sum_{\substack{\vec{v} \in \{0, 1\}^n\\v_{i_1} = 1,\\ v_{i_2} = \cdots = v_{i_d} = 0}}p_{\bm s}(\vec{v})= \frac{2+\Delta}{t}.
\end{align}
Additionally, to ensure that no test is empty, we set $P(\bm{s} = \vec{0}) = 0$. 
Conditions ~\eqref{cond: sum to one},~\eqref{cond: all positive}, and~\eqref{cond: dd whp} reduce our problem to proving the existence of non-negative solutions to a linear system.
To simplify our analysis, we restrict to the case where rows whose Hamming weights are equal have equal probability of being generated. For instance, if $n=3$, then we let $p_{\bm s}(100)=p_{\bm s}(010)=p_{\bm s}(001)$ and $p_{\bm s}(110)=p_{\bm s}(101)=p_{\bm s}(011)$.
This allows us to write our first conditions \eqref{cond: sum to one} and \eqref{cond: dd whp} as linear combinations of the probabilities that $\bm{s}$ has hamming weight $w \le n$, rather than the probabilities of $p_{\bm s}(\vec{v})$ for each $\vec{v} \in \{0, 1\}^n$. Let $\vec{p}$ be the vector whose $w^\text{th}$ element corresponds to the probability that $\vec{s}$ has hamming weight $w$. Conditions \eqref{cond: sum to one} and \eqref{cond: dd whp} can then be written as the following linear system, for which $\vec{p}$ is a solution
\begin{equation}
\label{eq: linsystem}
    \begin{bmatrix}
        \binom{n}{1} & \binom{n}{2} & \cdots & \binom{n}{n-d+1} & \;\;\;\cdots & \binom{n}{n}\\[.5em]
        \binom{n-d}{0} & \binom{n-d}{1} & \cdots & \binom{n-d}{n-d} & 0\;\;\cdots & 0
    \end{bmatrix}\vec{p} = \begin{bmatrix}
        1\\\frac{2+\Delta}{t}
    \end{bmatrix}.
\end{equation}
The first row of the matrix in equation \eqref{eq: linsystem} corresponds to the condition given by \eqref{cond: sum to one}; its $w^\text{th}$ component is the number of length $n$ binary vectors with weight $w$. The second row corresponds to the condition given by \eqref{cond: dd whp}; its $w^\text{th}$ component is the number of weight $w$ length $n$ binary vectors $\bm{s}$ with $\bm{s}_{i_1} = 1$ and $\bm{s}_{i_2} = \ldots = \bm{s}_{i_d} = 0$ for any distinct $i_1, \ldots, i_d \in [n]$.

The following lemma provides conditions on the parameters such that an elementwise nonnegative solution to the linear system~\eqref{eq: linsystem} exists.

\begin{lemma}
\label{thm:construction}
    For a general pool of \(n\) patients with a maximum of \(d\) infected, and any $\Delta > 0$, if \[\frac{\binom{n-d}{\lfloor \frac{n}{d} \rfloor - 1}}{\binom{n}{\lfloor\frac{n}{d}\rfloor}} > \frac{2+\Delta}{t},\] then equation \eqref{eq: linsystem} has an elementwise nonnegative solution. 
\end{lemma}

 We will use Farkas' Lemma to establish Lemma~\ref{thm:construction}.
    \begin{lemma}[Farkas' Lemma,~{\cite[p.~263]{Boyd--Vandenberghe2004}}]
        Let \(A \in \R^{m\times n}\) and \(\vec{b} \in \R^m\). Then exactly one of the following is true:
        \begin{enumerate}
            \item There exists \(\vec{p} \in \R^n\) such that \(A\vec{p} = \vec{b}\) and \(\vec{p} \ge 0\) elementwise
            \item There exists \(\vec{y} \in \R^m\) such that \(A^\intercal\vec{y} \ge 0\) elementwise and \(\vec{b}^\intercal\vec{y} < 0\)
        \end{enumerate}
    \end{lemma}
    
\begin{IEEEproof}[\bf Proof of Lemma~\ref{thm:construction}]
If no elementwise nonnegative vector \(\vec{p}\) exists satisfying \eqref{eq: linsystem}, then by the second condition of Farkas' lemma, there exists some \(\vec{y} = [y_1 \; y_2]^\intercal\) such that $y_1 \ge 0$, $y_1 + \frac{2+\Delta}{t}y_2 < 0$, and \(\binom{n}{w}y_1 + \binom{n-d}{w-1}y_2 \ge 0\) for all \(w = 1, 2, \ldots, n-d+1\).
Together, these imply 
    \begin{align*}
        0 &> y_1 + \frac{2+\Delta}{t}y_2\\
        &\ge \frac{-\binom{n-d}{w-1}y_2}{\binom{n}{w}} + \frac{2+\Delta}{t}y_2\\
        &= y_2\left(\frac{2+\Delta}{t} - \frac{\binom{n-d}{w-1}}{\binom{n}{w}}\right),
    \end{align*}
for all $w = 1, 2, \ldots, n-d+1$. Note that, since $y_1 \ge 0$ and $y_1 + \frac{2+\Delta}{t}y_2 < 0$, we must have $y_2 < 0$. This means
$$\frac{\binom{n-d}{w-1}}{\binom{n}{w}}< \frac{2+\Delta}{t}.$$
We compute the maximum of $f(w) \triangleq \frac{\binom{n-d}{w-1}}{\binom{n}{w}}$ over $w \in [d]$. Note that
\begin{align*}
        f(w) &= \frac{w\prod_{i=0}^{w-2}(n-d-i)}{\prod_{i=0}^{w-1}(n-i)},
\end{align*}
so that 
\begin{align*}
        f(w+1) &= \frac{(w+1)(n-d-w+1)\prod_{i=0}^{w-2}(n-d-i)}{(n-w)\prod_{i=0}^{w-1}(n-i)}.
\end{align*}
We examine the terms outside of the products in the numerator and denominator, which differ between $f(w)$ and $f(w+1)$. For $f(w+1)$, we have
\begin{align*}
    \frac{(w+1)(n-d-w+1)}{n-w} &= w\left(1 + \frac1w\right)\left(1-\frac{d-1}{n-w}\right).
\end{align*}
If $w \ge \lfloor n/d \rfloor$ is an integer, then $wd \ge n$ so that 
\begin{align*}
    w(d-1) &\ge n-w\\
    \frac{d-1}{n-w} &\ge \frac{1}{w},
\end{align*}
and $f(w+1) \le f(w)$. If $w < \lfloor n/d \rfloor$ is an integer, then $w \le \frac{n}{d}-1$ so that 
\begin{align*}
    n-w &\ge d(w+1)-w\\
    &> (d-1)(w+1)\\
    \frac{1}{w} &> \left(\frac{d-1}{n-w}\right)\left(1+\frac1w\right),
\end{align*}
and $f(w+1) > f(w)$. As such, $f$ is maximized when $w = \lfloor n/d \rfloor$. Hence, if \[\frac{\binom{n-d}{\lfloor \frac{n}{d} \rfloor - 1}}{\binom{n}{\lfloor\frac{n}{d}\rfloor}} > \frac{2+\Delta}{t},\] then there exists some elementwise non-negative vector \(\vec{p}\) satisfying \eqref{eq: linsystem}. Thus, there exists a valid probability distribution on $\mathbf{s}$, using which we can generate a $(t, n, d)$-tropical code with high probability.
\end{IEEEproof}

\subsection{Performance of the probabilistic construction}
\label{sec:performance}
In this section, we establish conditions for which the proposed construction provides us with a valid tropical code.
\begin{thm}
\label{thm:performance}
    The construction given in Section \ref{sec:construction} is a $\left(t, n, d, \epsilon\right)$-tropical code if
    $$t > 2e\left(d^2\log\left(\frac{ne}{d}\right) + d\log\left(\frac{d}{\epsilon}\right) + 2d\right).$$
\end{thm}
We will use Chernoff bound to establish Theorem~\ref{thm:performance}.
\begin{lemma}[Chernoff bound~\cite{Mitzenmacher--Upfal2017}]
\label{lemma:chernoff}
    Let $X = \sum_{k=1}^t X_k$ where all $X_k\sim\bern(p)$ are independent. Let $\mu = \mathbb{E}(X) = tp$. Then
    $$\P(X \le (1-\delta)\mu) \le e^{-\mu\delta^2/2},$$
    for all $0 < \delta < 1.$
\end{lemma}

\begin{IEEEproof}[\bf Proof of Theorem~\ref{thm:performance}]
For a given $\epsilon \in (0, 1)$, choose $\Delta= \left(2+2\log\left(\frac{d}{\epsilon}\binom{n}{d}\right)\right)$. For some $d$ fixed distinct blocks indexed $i_1, \ldots, i_d \in [n]$ let $X_k = \mathbf{1}_{\{{\bm S}_{k, i_1} = 1, {\bm s}_{k, i_2} = 0, \ldots {\bm s}_{k, i_d} = 0\}} \sim \bern(\frac{4+2\log\left(\frac{d}{\epsilon}\binom{n}{d}\right)}{t})$ be the random variable indicating whether or not the $i_1^\text{th}$ entry of the $k^\text{th}$ row is $1$ and the $i_2, \ldots, i_d$ entries of the $k^\text{th}$ row are 0. Then the probability of error for the chosen subset of blocks is $\P\left(\sum_{k\in[t]} X_k < 2\right)$, since the sum $\sum_{k\in[t]} X_k$ gives the size of the set $\left|B_{i_1}\setminus\left(\bigcup_{j=2}^dB_{i_j}\right)\right|$. Then note that 
\begin{align*}
    &\P\Big(\sum_{k\in[t]} X_k < 2\Big) \\
    &= \P\left(\sum_{k\in [t]} X_k < \Big(4+2\log\Big(\tfrac{d}{\epsilon}\tbinom{n}{d}\Big)\Big)\left(1 - \tfrac{2+2\log\left(\frac{d}{\epsilon}\binom{n}{d}\right)}{4+2\log\left(\frac{d}{\epsilon}\binom{n}{d}\right)}\right)\right).
\end{align*}
We may then use the Chernoff bound (lemma~\ref{lemma:chernoff}) with $\mu = 4+2\log\left(\frac{d}{\epsilon}\binom{n}{d}\right)$ and $\delta = \frac{2+2\log\left(\frac{d}{\epsilon}\binom{n}{d}\right)}{4+2\log\left(\frac{d}{\epsilon}\binom{n}{d}\right)}$ to get:
\begin{align*}
    &\P\Big(\sum_{k\in [t]}X_k<2\Big) \\
    &\le \exp\left(-\left(4+2\log\left(\tfrac{d}{\epsilon}\tbinom{n}{d}\right)\right)\tfrac{\left(2+2\log\left(\tfrac{d}{\epsilon}\binom{n}{d}\right)\right)^2}{2\left(4+2\log\left(\tfrac{d}{\epsilon}\binom{n}{d}\right)\right)^2}\right)\\
    &= \exp\left(-\frac{\left(2+2\log\left(\frac{d}{\epsilon}\binom{n}{d}\right)\right)^2}{2\left(4+2\log\left(\frac{d}{\epsilon}\binom{n}{d}\right)\right)}\right).\\
\end{align*}
Now we take the union bound over every subset of $[n]$ with cardinality $d$ with one specific index chosen. This gives an upper bound for $P_e^{n, d}$.
\begin{align*}
    P_e^{n, d} &\le \sum_{\substack{I\subset[n], |I| = d\\i_1\in I}}\P\left(\sum_{k \in [t]}X_k<2\right) \\ 
    &\le d\binom{n}{d}\exp\left(-\frac{\left(2+2\log\left(\frac{d}{\epsilon}\binom{n}{d}\right)\right)^2}{2\left(4+2\log\left(\frac{d}{\epsilon}\binom{n}{d}\right)\right)}\right)\\
    &\le d\tbinom{n}{d}\exp\left(-1-\log\left(\tfrac{d}{\epsilon}\tbinom{n}{d}\right)\right)\\
    &\le d\tbinom{n}{d}\exp\Big(\log\Big(\tfrac{\epsilon}{d\tbinom{n}{d}}\Big)\Big)\\
    &=\epsilon
\end{align*}

To study the asymptotic behaviour of the construction in Lemma~\ref{thm:construction}, we reduce our final expression using the following asymptotic approximation of the binomial coefficient $$\binom{n}{\alpha n}\approx\frac{1}{\sqrt{2\pi\alpha(1-\alpha)n}}2^{nh(\alpha)},$$
where $h(\cdot)$ is the binary entropy function. This gives 
\begin{align*}
    \frac{\binom{n}{\lfloor\frac{n}{d}\rfloor}}{\binom{n-d}{\lfloor \frac{n}{d} \rfloor - 1}} \approx \sqrt{\frac{n-d}{n}}2^{dh(1/d)}
    < \sqrt{\frac{n-d}{n}}ed,
\end{align*}
so that our construction is a $(t, n, d, \epsilon)$-tropical code if
\begin{equation}
\label{eq:grossbound}
    t > \sqrt{\frac{n-d}{n}}ed\left(4+2\log\left(\frac{d}{\epsilon}\binom{n}{d}\right)\right).
\end{equation}
Using the bounds $\sqrt{\frac{n-d}{n}} < 1$ and $\binom{n}{d} \le \left(\frac{en}{d}\right)^d$ we can relax the condition in~\eqref{eq:grossbound} so that our construction is a $(t, n, d, \epsilon)$-tropical code if
\begin{equation}
    t > 2e\left(d^2\log\left(\frac{ne}{d}\right) + d\log\left(\frac{d}{\epsilon}\right) + 2d\right).
\end{equation}
\end{IEEEproof} 

\section{Comparison}
Since both probabilistic and deterministic constructions exist for disjunct matrices, and our result in Theorem~\ref{thm:code from doubly disjunct} provides a construction for a $(d-1)$-doubly disjunct block design from a $(d-1)$-disjunct block design, we have three ways to produce $(t, n, d, \epsilon)$-tropical codes for some fixed $d$:
\begin{enumerate}
    \item First construct a deterministic $(d-1)$-disjunct matrix, then use Theorem~\ref{thm:disjunct to doubly disjunct} to construct a $(d-1)$-doubly disjunct matrix;
    \item Construct a $(d-1)$-disjunct matrix probabilistically, then use Theorem~\ref{thm:disjunct to doubly disjunct} to construct a $(d-1)$-doubly disjunct matrix;
    \item Construct a $(d-1)$-doubly disjunct random matrix directly using Lemma~\ref{thm:construction}.
\end{enumerate}
Theorem~\ref{thm:disjunct to doubly disjunct} doubles the number of tests used, but does not change the asymptotic behaviour of the construction used for a $(d-1)$-disjunct matrix. The deterministic constructions given by~\cite{Porat--Rothschild2007,Kautz--Singleton1964} have $t = O(d^2\log n)$ with $\epsilon = 0$ for specific parameter choices. As such, the first method also gives $t = O(d^2\log n)$ with $\epsilon = 0$. 
For the second method, we can use our construction detailed in Section~\ref{sec:construction} to construct $(d-1)$-disjunct matrices by replacing the right hand side of Lemma~\ref{thm:construction} and all subsequent calculations with $\frac{1+\Delta}{t}$. This gives $(d-1)$-disjunct matrices with $t > e\left(2d^2\log\left(\frac{ne}{d}\right) + 2d\log\left(\frac{d}{\epsilon}\right) + 3d\right)$ for error bound $\epsilon$. Then, using Theorem~\ref{thm:disjunct to doubly disjunct}, we can get a $(d-1)$-doubly disjunct matrix with $t > 2e\left(2d^2\log\left(\frac{ne}{d}\right) + 2d\log\left(\frac{d}{\epsilon}\right) + 3d\right)$, which is greater than the bound in Theorem~\ref{thm:performance} for the construction of a $(d-1)$-doubly disjunct matrix directly, as in the third method. From equation~\eqref{eq:grossbound} in Theorem~\ref{thm:performance}, the construction given in Section~\ref{sec:construction} requires $t = O\left(d\log\binom{n}{d} + d\log\left(\frac{d}{\epsilon}\right)\right)$. When $d = o(n)$ is sublinear, then we have $t = O(d^2\log n + d\log(d/\epsilon))$. If $d = \Theta(n)$, then $t = o(d^2\log n + d\log(d/\epsilon))$.

In a practical setting, if the situation requires zero error, then we can use the first method to construct a doubly disjunct matrix with the set of given parameters. Otherwise, if some error $\epsilon$ is tolerable, then our construction in Section~\ref{sec:construction} may be preferable.

\section{Conclusion}
\label{sec:conclusion}

In this paper we consider a probabilistic construction for a $(d-1)$-doubly disjunct block design, as well as a deterministic method based on the existence of a $(d-1)$-disjunct block design. We show that constructing a $(d-1)$-doubly-disjunct block design probabilistically gives a similar performance to a fully deterministic construction based on a $(d-1)$-disjunct block design, with the added benefit of being able to use arbitrary parameters.
\balance
\clearpage


\bibliographystyle{IEEEtran}
\bibliography{references}



\end{document}